%% file: MBT-2012.tex
\documentclass[submission,copyright]{eptcs}
\providecommand{\event}{MBT 2012} 
\usepackage{breakurl}
\usepackage{graphicx}
\usepackage{fancyvrb}
\usepackage{tabularx}
\newcolumntype{L}[1]{>{\raggedright\arraybackslash}p{#1}}
\newcolumntype{C}[1]{>{\centering\arraybackslash}p{#1}}
\newcolumntype{R}[1]{>{\raggedleft\arraybackslash}p{#1}}

\title{Rule-based Test Generation with Mind Maps}
\author{Dimitry Polivaev
\institute{Giesecke \& Devrient GmbH\\ Munich, Germany}
\email{dimitry.polivaev@gi-de.com}
}
\def\titlerunning{Rule-based Test Generation}
\def\authorrunning{Dimitry Polivaev}
\begin{document}
\maketitle

\begin{abstract}
This paper introduces basic concepts of rule based test generation with mind maps, 
and reports experiences learned from industrial application of this technique in the domain of smart card testing
by Giesecke \&Devrient GmbH over the last years. It describes the formalization of test selection criteria used by our test generator,  our test generation architecture and test generation framework.
\end{abstract}
\input{mbt2012-body}
\nocite{*}
\bibliographystyle{eptcs}
\bibliography{references}
\end{document}

%% file: mbt2012-body.tex
\section{Introduction}
\par Testing is very important for smart card development because failure costs can be very high. Smart card
manufacturer Giesecke \& Devrient GmbH spends significant effort on implementation of tests and the development of
new testing techniques. This paper presents some of the latest methodological results.
\par Our test cases should be short and easy to understand. They should execute fast and run independently from
other test cases, which simplifies debugging in case of test failure. In the past all test cases were
separately specified and implemented by test engineers. For testing of complex systems thousands of such test
cases containing setup, focus, verification and post processing were implemented. Manual development of
redundant code of the test scripts resulted in high costs of maintenance and change. We studied existing test
generation techniques and decided to develop our own approach.
\par Known test generation techniques often concentrate on only one modeling abstraction. For instance the
classification tree method \cite{CTE} provides exploration of test case space based on systematic partitioning
of test input and output data. It uses classification tree as a model of a test space. It does not make use of
SUT models e.g. for automatic data partitioning or selection of relevant data combinations.
\par There are also classical model based approaches where test cases and test coverage are generated from the
SUT models. This technique is supported by different test generation tools like Conformiq Designer
\cite{Conformiq}, LEIRIOS Test Designer \cite{Leirios} or Smartesting CertifyIt \cite{Smarttesting}. They do
not consider test strategies to be the responsibility of test engineers. The test strategies are implemented in
the tools. To influence the test generation the SUT model should be changed, other ways to introduce e.g.
additional variations of test data or different preconditions leading to the same model state are not
supported. Hence the models become more complex, they include test specific elements and can not be reused for
different kinds of tests.
\par Also, models usually need to be complete, i.e. they need to contain all actions which can be used by test
cases. As a consequence, for complex systems where test preconditions need to refer to different subsystems, the
latter must be modeled as well because there is no other way to add the required actions to the generated test
cases. And finally, test case generation using model state exploration techniques have high requirements on
computer power and limitations on model state space.
\par Usage of different kinds of models allows software engineering principles to be applied to test
development. Identification of different abstractions makes it possible to use replaceable components for
controlling different aspects of the test generation. An interesting example of separating different aspects
of test generation in different models is given by Microsoft Spec Explorer \cite{Specexplorer}. There are
models of SUT and also so-called slices, representing test scenarios considered by the generator. However it
still focuses on model state exploration making the SUT model the central element of test generation with the
limitations described above.
\par Some other tools model test sequences. This technique is represented by e.g. .getmore \cite{MZT}. Here a
test sequence model is mixed with a SUT model, making it hard to reuse them separately. They also do not care
about the specification of different data inputs within the same test sequence or definition of model independent
test coverage goals.
\par In our approach there are three kinds of models. The most important one is a {\bf test strategy.} It
models a test case space and is responsible for the amount and variance of the generated test cases. Test
strategies are placed in the focus of the test development. Test engineers explicitly define them, taking
advantage of their knowledge, intuition and experience. Then there is a {\bf model of system under test (SUT)}
which simulates SUT data and behavior and, last nut not least, a {\bf test goal model} which defines the test
coverage. All of these model types operate with different abstractions and need different notations.
\par This paper explains a formalization of test strategy and test goals used by our test generator. It also
describes test generator architecture and a test generation framework. The framework allows generation of
test suites from modular reusable components. In addition to the three kinds of models listed above there are
components called solvers and writers responsible for the output of the test scripts.
\par {\bf Structure of this paper.} This paper describes models and other components used in our test
generator. Section 2 introduces an example SUT used in all following sections to demonstrate the introduced
concepts. Property based test strategies developed as rule sets are explained in section 3. Test goals used for
test selection based on test coverage measurement are introduced in section 4. Section 5 describes the other
test generation components and demonstrates our test generation architecture. Our experiences from
the card development projects using this test generation technique are then reported. The paper ends with a small
conclusion section.
\section{Example: calculation of phone call costs}
Let us consider the following specification as an example of a system under test. It specifies software for the
calculation of phone call costs. The interface consists of two functions:
    \begin{itemize}
  \item 
        setCheapCallOptionActive(isActive)
      \item 
        calculateCallPrice(country, phoneNumber, day, callBeginTime, callDuration)
      
    \end{itemize}
      The first function is used to activate or deactivate a special tariff.
\par Function calculateCallPrice satisfies the following requirements:
\par If an invalid country or phone number is used, the function returns 0, otherwise the used tariff is given
by a price table:
\begin{table}[h]
\begin{tabular}{|C{2.5cm}|C{2.5cm}|C{2.5cm}|C{2.5cm}|C{2.5cm}|}
\hline
{\bf 
          Destination
        } & {\bf 
          Standard \newline
Tariff
        } & {\bf 
          CheapCall \newline
Tariff
        } & {\bf 
          At night, \newline
on the weekend
        } & {\bf 
          Time unit \newline
(seconds)
        } \\
\hline
          National
         & 
          0.10\$
         & 
          0.07\$
         & 
          0.03\$
         & 
          1
         \\
\hline
          International\_1
         & 
          1.00\$
         & 
          0.50\$
         & 
          0.80\$
         & 
          20
         \\
\hline
          International\_2
         & 
          2.00\$
         & 
          1.20\$
         & 
          1.80\$
         & 
          30
         \\
\hline
\end{tabular}
\caption{Call costs pro time unit dependent on call time and destination.}
\end{table}
      The destination can be obtained from the country as follows:
    \begin{itemize}
  \item 
        if country is empty or 'National' the destination is set to National ,
      \item 
        if country is 'Greenland', 'Blueland' or 'Neverland' the destination is set to International 1,
      \item 
        if country is 'Yellowland' or 'Redland' the destination is set to International 2
    \end{itemize}
\par All other country values are invalid.
\par Tariff "At night" applies if the call begins between 8 pm and 6 am.
\par Tariff "On the weekend" applies if the call begins on Saturday or on Sunday.
\par Call duration given in seconds is rounded up to units specified in column "Time unit".
\par Maximal call duration is limited to 24 hours.

\section{Exploration of test case space with test strategies}
\par Tests are developed as a part of the quality assurance process. They should check that software implementation satisfies the user requirements. Each tested requirement should be covered by a sufficient set of test cases checking all relevant aspects. There is an infinite number of possible test cases many of which should be implemented. Hence test development requires good systematics to choose which ones to actually implement.
    
\par We call a model, describing how test case space is systematically explored, a {\bf test strategy}. It
defines the number and variance of test cases in a generated test suite. For example, the test strategy could define which combinations of valid and invalid input data should be used. As we explain later, test strategies usually not only guide input data coverage, they also describe other test aspects e.g. test intention (e.g. a good case or a bad case),  expected results or a test case name. 
\par The base concept of test strategy definitions is a test case property.
  \subsection{Test case properties}
The {\bf test case property} is a key-value pair. Each property describes a test case characteristic such as test
name, variations of test actions or their input data, other test case related data like expected results, test
coverage and so on. Test properties can be used to declare what features are tested by a particular test case
or to classify tests in a test suite in some another way. The keys and values can be represented by objects of
arbitrary type. In our implementation the keys are strings and the values are some hierarchical data
structures.
\par Some of the test case property values e.g. some of the input parameters can be assigned independently, but
there may also be dependencies between property values. For example, in addition to properties corresponding to
input parameters of the method "calculateCallPrice", the test strategy could define classifying properties
    \begin{itemize}
  \item 
        i{\it sCallValid}  with possible values {\it true} and {\it false},
      \item 
{\it failureReason} with possible values {\it invalidCountry} and {\it invalidNumber}, which makes sense only
if the call is not valid.
      \item 
        {\it destination} with values {\it National}, {\it International\_1}and {\it International\_2.}
      
    \end{itemize}
According to the requirements, if test case property {\it destination} is assigned value {\it
International\_1}, parameter {\it country} can be chosen only from values {\it Greenland}, {\it Blueland} or
{\it Neverland.}
\par A test strategy specifies range and number of the generated test cases by describing combinations of the
test case property values considering their dependencies. The test case space covered by a test strategy is
completely defined by the strategy property space. When all required property values are set, a script with the
test case can be written.
\subsection{Rule-based exploration of property space}
\par Test strategies can be represented by a collection of business rules building an ordered {\bf rule set}.
The rules specify variations of the test property values and dependencies between them.
\par Each rule defines values of one property called its {\bf target property}. Because each test case definition depends on values of many properties, the test strategy is given by a set of rules. The values can be defined as
expressions referencing other property values and arbitrary function calls. The rule can also add new rules to
the rule set. The added rules have other target properties. They remain in the rule set, as long as iterations
over the values of the target property of the rule where they are defined continue.
\par There are two kinds or rules: {\bf iteration rules} and {\bf default rules}.
\par The iteration rules define lists of values for their target properties. Their evaluation is triggered by
assignment of a value to another property set by another iteration rule (so-called forward chaining).
\par The default rules specify values of properties that have not been assigned by the iteration rules. Such a
rule can be evaluated when its target property value is requested by another rule or by the test generation
algorithm (so-called backward chaining).
\subsubsection{Iteration rules}
\par Each iteration rule contains three parts: the {\bf WHEN}-part describes when the rule is applied, the
optional {\bf IF}-part describes an additional condition and the {\bf THEN}-part describes the action setting
the list of property values and sometimes adding new rules to the set.
\par The WHEN-part can reference an arbitrary number of properties. If it does not reference any property,
the rule is applied when test generation starts, otherwise it waits until all referenced properties become
assigned by actions of other iteration rules.
\par If the rule is executed and its condition is satisfied, then its target property is sequentially assigned
values from the value list set by the rule action. Each assignment triggers evaluation of the dependent
iteration rules. Each test generated by the rule engine can be uniquely identified by the values of the
properties iterating over more than one value.
\par The iteration value list can be attributed with an option "shuffled" changing the order of the list
elements every time the iteration resumes from the beginning of the list. It produces different random value combinations
of parallel iterated properties. Hence if generation runs with different seeds it creates different test cases.
\par If a rule sets an empty list of property values, it means that no value can be found for the already set
values of other properties. In this case rule engine tries to change a value of some property referenced in
its WHEN part.
 \par \textbf{Some informal examples of the iteration rules.}
\par Let us look at some examples of the iteration rules. They will be repeated in mind map notation in a later
part of this article.
\par Iteration rule 1: {\bf WHEN} (empty) {\bf IF} (empty) {\bf THEN} property {\it isCallValid} is
sequentially assigned values {\it true}, {\it false}
\par Iteration rule 2: {\bf WHEN} property {\it isCallValid} is assigned {\bf IF} it has value {\it true }{\bf
THEN} property {\it destination} is sequentially assigned values {\it National, International\_1} and {\it
International\_2}
\par Iteration rule 3: {\bf WHEN} property {\it isCallValid} is assigned {\bf IF} it has value {\it true }{\bf
THEN} property {\it callDuration} is sequentially assigned values {\it 1} and {\it 60}
\par Iteration rule 4: {\bf WHEN} property {\it destination} is assigned {\bf IF} it has value {\it
International\_1} {\bf THEN} property {\it country} is sequentially assigned values {\it Greenland, Blueland,
Neverland}

\subsubsection{Default rules}
\par Default rules contain only an optional IF - part with a condition and a THEN - part describing an action.
They are evaluated whenever their target property value is requested from an IF- or THEN-part of another rule
or from any other test generation component but has not been assigned yet. The action assigns some value to the
rule's target property. Default rules are not designed for producing iterations, they always assign single
values.\par \textbf{Informal example of a default rule}
\par Default rule 1: {\bf IF} {\it destination} = {\it International\_2 }{\bf THEN} assign to {\it country}
value {\it Redland}
 \subsubsection{Rule stack}

\par Iteration rules with the same properties in the WHEN part and the same target property build a so-called
rule stack. They are processed in the opposite order to their definition. If some rule with empty condition or with
satisfied condition was found no other rules from the stack are executed. There are also rule stacks with
default rules.
\par Rule sets must be self-consistent: if a property has already been assigned, no other iteration rule may
try to reassign it.
\par Rules added to the stack later can override previously defined rules. This can be used to define strategy
variations overloading some default strategy. For example there can be some common and product dependent
subsets for testing of a product line.
\subsubsection{Rule engine}
\par There is a rule engine generating combinations of test case properties corresponding to single test cases.
It starts with a given rule set and executes all iteration rules from rule stacks with an empty WHEN part,
iterates over all values from the property lists and processes chained iteration rules as properties become
assigned. Each combination of property values can become a test case.
\par The rule engine tracks all dependencies between properties. Property A is called dependent on property B
if there is a rule with target property A which refers to property B from its WHEN, IF or THEN part. When a
property iterates over different values, the next value is assigned only after all iterations over its dependent
properties finishes.
\par The rule engine run terminates after all iterations specified by iteration rules are finished.
\par Whenever an unassigned property is requested, the rule engine executes a related default rule if it is
available and applicable according to its IF-part.
\subsubsection{Example}
\par Iteration rules 1 to 4 from the above example generate following property combinations:
\par \begin{small}
\begin{Verbatim}[frame=single]
1:$isCallValid:TRUE/$destination:National/$callDuration:1
2_$isCallValid:TRUE/$destination:International_1/$country:Greenland/$callDuration:60
3:$isCallValid:TRUE/$destination:International_1/$country:Blueland/$callDuration:1
4:$isCallValid:TRUE/$destination:International_1/$country:Neverland/$callDuration:60
5:$isCallValid:TRUE/$destination:International_2/$callDuration:1
6:$isCallValid:FALSE
\end{Verbatim}
\end{small}
\par The rule engine simultaneously iterates over values of properties {\it country}  and {\it callDuration}. If their lists
had the option {\it shuffled} set, mutual combinations of their values would be randomized.

\subsubsection{Mind map representation}
\par The rule sets modeling test case space can be implemented in scripts or with mind maps.
\par In the mind maps relations between properties are represented by relative positions of the mind map nodes.\par All examples given in figures 1,2 and 3 demonstrate how the above set of rules can be implemented in a
mind map.\begin{figure}[htb]
\begin{center}
\includegraphics[width=12cm]{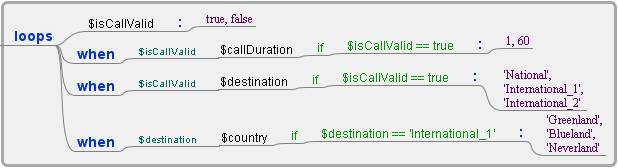}
\caption{Exact copy of the above rules in the mind map notation.}
\end{center}
\end{figure}
\begin{figure}[htb]
\begin{center}
\includegraphics[width=12cm]{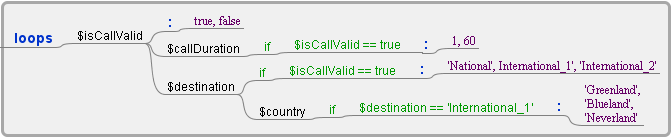}
\caption{Here the WHEN relation is represented by a property definition outgoing from another property
definition. The IF-parts are specified explicitly.}
\end{center}
\end{figure}
\begin{figure}[htb]
\begin{center}
\includegraphics[width=12cm]{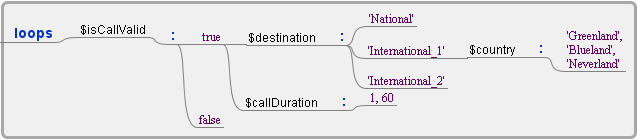}
\caption{The map structure is used for representing both WHEN and IF parts. IF equality conditions are
specified by property definitions outgoing from value definitions.}
\end{center}
\end{figure}

\par Use of mind maps for test strategy implementation simplifies development, reviewing and improvement of the
strategy. Mind maps as a representation of generation rules offers good visualization, automatic context
dependent node formatting, search and filtering of the rule sets.
\section{Test coverage controlled by test goals}
\par All possible combinations of input parameters cannot be tested within a reasonable amount of
time. If test strategies generate too many test cases, the test cases contributing to the achievement of desired coverage criteria must be distinguished from the test
cases, which can be discarded without negative effects on coverage. Therefore test selection and requirement
coverage criteria should be defined in addition to the test strategy rules. Such criteria are called {\bf test
goals}.
\par Test goals can also be used to check completeness and correctness of test generation. If some goals are
not achieved, it can indicate an error in test generation components or in the specification.
\par We differentiate between finite and infinite test goals. A {\bf finite test goal} is basically a check
list which can contain SUT model code coverage, the modified condition/decision coverage or the  boundary condition coverage, values of the test input parameters, model predicted results or model interim data separately or in a combination. For instance, in our phone call example, a test goal can require that some calculated call prices are covered by the complete test suite or by its subset, limited to calls to some particular destination. 
\par The finite goal is defined as a pair of the complete check list given as e.g. set of strings and a function mapping test case
data to the set of values containing the check list values. A test is important for a finite goal, if
the goal function called with the test related data returns some value contained in the check list for the
first time since the generation was started.
\par Because sometimes calculation of the complete check list is difficult or even impossible as in the case
of model path coverage, there are also so-called {\bf infinite goals}. These consist only of their goal
function and there is no predefined check list. The test is important for an infinite goal if the function returns
a result not previously returned. The check list is automatically filled with all returned values.
\par The test goal functions can be defined as expressions using the test case property values. And every expression defined on test case property values can be used to define a new test case property itself. For instance the goal functions can use SUT model execution results, its intermediate states,  collected code coverage and state coverage data. A set of all statements contained in the model can be obtained from the model automatically. It builds a check list for a goal based on the model code coverage. Check lists for state coverage or result coverage related goals can be separately created by test engineer. This way coverage of the test suite based on model code and state coverage is ensured. 
\par Given a set of goals, test generation statistics are collected during the test generation. They describe
how often any particular goal value was returned for the generated test cases.
\par Goals related to combinations of test case properties specified in the strategy rules can be defined in the mind maps
as in our example. The example test goal given in figure 4, requires that all possible combinations
of country and tariff are tested with a valid call.\begin{figure}[htb]
\begin{center}
\includegraphics[width=12cm]{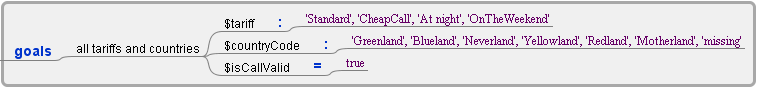}
\caption{Example of property based test goal definition in a mind map}
\end{center}
\end{figure}
\section{Test engineering}
\par For the sake of maintainability and changeability of test
suites they should be generated from modular organized reusable
components. The complete test generation architecture is
described in this section.
\subsection{Test components}
\par Previous sections explained {\bf test strategies} and {\bf test goals} used for test variant generation,
selection and test coverage estimation. These, together with a so-called test script {\bf writer,} call
functions from the SUT related components {\bf model} and {\bf solver} which model the behavior of system under
test and know, or are able to calculate how the SUT state can be manipulated or checked.
\par The {\bf writer} is a component which creates executable test scripts from the
test property values. For instance it outputs test header, test name, description,
commands, comments etc. It is the only module which depends on the language and libraries used by the generated
test scripts. It is called by a framework for each generated test case separately.
\par The {\bf model} calculates expected results and reports data for statement coverage, path coverage, modified condition /
decision coverage, internal states at given statements, given a function name, input parameters and data state
before call. The collected coverage information can be used for test coverage measurement and for test goal
definition. A model code coverage related test goal can be defined as a pair of the set of all statements
contained in the model and a function returning a list of statements covered by the given test case. Similar
goals can be defined for measuring of the modified condition / decision coverage and of boundary coverage.
\par The {\bf solver} decides which SUT functions should be called and which parameter values should be used in
order to bring the SUT into a required state and to check its state. For instance, test commands in preconditions, post processing or
verification can be calculated in this component. In our example there is a function
setCheapCallOptionActive(isActive). To use this function in a precondition the writer can obtain the complete precondition data
from the solver.
\par The solver can be implemented using tools statically analyzing the model. However direct implementation is often easier and more efficient.  
It is particularly the case for complex systems, where test preconditions require calls to different subsystems not covered by the model.
\par If a suitable solver is available, the test strategy can directly define iterations over test output data related or model state
related test case properties. It is useful for generating a test suite with test cases covering wished output data or
SUT model states. Implementing such solvers can be difficult. Thus test strategies more often iterate over different
preconditions and test input data. Then the output data, code and state coverage are controlled by the test goals and
not by the strategy. This approach makes complex solvers superfluous but requires much more test case property combinations to be tried 
before all goals are achieved. 
\begin{figure}[htb]
\begin{center}
\includegraphics[width=14cm]{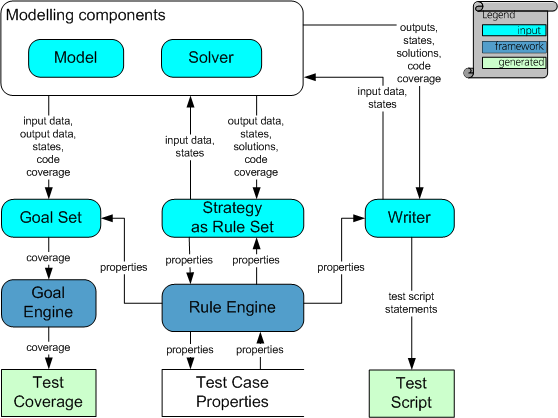}
\caption{Test generation data flow}
\end{center}
\end{figure}
\subsection{Component reusability}
\par The proposed design simplifies reuse of the same components for different tasks because they can be
replaced independently in order to create different test suites.
  \begin{itemize}
  \item 
        Different strategies can be used for testing of different features of SUT.
      \item 
        Different models can be used for testing of different products / product variations.
      \item 
Different solvers can be used to prepare an initial state of the test target, to verify its
state after executing the focus or reset it at the end of the test case using different input data and command sequences.
      \item 
        Different writers can be used for generating scripts for different script languages.
      \item 
Different test goals can be used for variation of test depth, e.g. for smoke tests, regression tests,
exhaustive tests etc.
    \end{itemize}

 \subsection{Test generation framework}
\par The described ideas have been implemented as a test generation framework. The framework allows modular
development of test suites similar to development of software. The test components are comparable to software
components in a software development. The test generator produces test scripts and test coverage information
like a compiler building executable software applications.\begin{figure}[htb]
\begin{center}
\includegraphics[width=10cm]{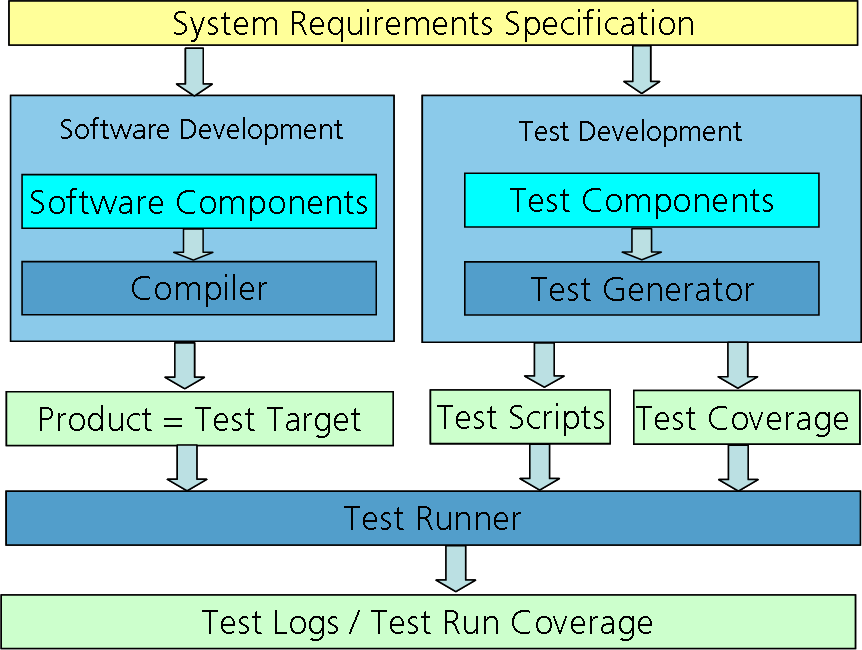}
\caption{Test generation process compared to software development}
\end{center}
\end{figure}
\par Our framework contains tools and libraries for the development of the test components and for the
generation of executable test scripts in arbitrary textual languages.
\par For the development of test strategies and test goals as mind maps, the framework uses mind map editor
Freeplane, which is available under GPL. It allows you to specify rules as mind maps, and supports mind map
development through context dependent text formatting, map filtering and search. There is a converter
transforming mind map based sets of rules and goals into the generation scripts.
\par The framework contains a rule engine evaluating the rules and a goal engine collecting information about
test coverage achieved by generated test cases. Using information from the goal engine, our rule engine can
also effectively reduce the explored property space, and thus the number of the generated property combinations, skipping
iterations over properties not relevant for the achievement of the test goals. This procedure is similar to the
one described in \cite{Korat}.
\par The framework contains a compiler for the programming language used for the development of writer and model
components. The compiler generates code for automatic measurement of code, path and mc/dc coverage and boundary
condition coverage needed for the coverage related goals. For the purpose of creating the test script files it
supports test script templates embedded into the generation code. Such templates can access script variables
and functions. The language has support for large integer arithmetic, complex data structures, it contains
closures and statement blocks as method arguments. The test generator software translates scripts written in
this language into java code, which is then compiled and executed. Java libraries can be called directly
from the scripts if it is required.

\section{Framework evaluation based on real life experience} 
\par The framework is currently used in projects
testing the newest MasterCard, VISA smart cards and java cards. It demonstrated excellent scalability. In big
projects it was used to generate tens of thousands of test cases but it was also efficient for generating small
test suites containing a few hundred of test cases. For example, the test strategy definition of one typical project
consists of 1135 iteration rules and 202 default rules, defined in 4007
nodes contained in 14 mind maps. The model implemented 1467 functional
requirements. The test goal definitions were based on the complete code coverage, the modified
condition/decision coverage and boundary coverage of the SUT model. They could be obtained automatically. 
The generator reduced the explored property space as described above, generated 90389 property value
combinations, selected and created 5312 test cases. 
\par Our overall positive experience has shown that the described methodology can be effectively used for test generation for different test targets. Test implementation
effort has been reduced by more than 50\% compared to methods not using the test generation. We achieved even greater
reduction of test adaptation efforts on specification updates because they often caused only minor changes of the
SUT model and the test generation could be repeated. 
\par Test development using the proposed methodology fits well
in existing processes. It is helpful to start with the manual implementation of some test cases which can later be used for
creating the test templates embedded in a writer component. After they are available, development of all necessary
components can be started. In a project team, they are implemented by different people at the same time, giving a
natural way to divide the work. All components can be developed incrementally. It is possible to have a small
number of test cases, with a high coverage of the requirements, available quickly. While the first found bugs are
being fixed, the strategies can be extended. And there is no need to implement a full model of the SUT. Only aspects
required to define a strategy and test goals or to calculate expected results should be considered. 
\par The method achieves better test case systematics compared to manual test development we used before. It results from the use of
formally defined test strategies and the monitoring of test coverage by test goals. Test case input data determined by
the test strategy with random values and mutually random value combinations for different properties additionally
increase test coverage and help to discover further bugs. Specification errors can be found earlier with the aid of
model code coverage information and test goals statistics collected during the test generation. Detailed debugging
information like the model code coverage is supplied. This also helps to reduce effort for test object debugging.

\section{Conclusion}
\par Our methodology identifies different kinds of test generation related models. It offers different
abstractions and notations for their definition and suggests to implement them in different ways, and just as
detailed as necessary to achieve the test goals.
\par It sees the test strategy as the driving force of the test generation and explains how to implement it
using a special form of business rules systematically defined in mind maps. The maps look similar to classification
trees known from the classification tree method but they are more powerful. Their strength is to consider
complex dependencies between property values, visually specify their relations and dynamically change a rule
set during the strategy run.
\par Information obtained from SUT models including model code coverage data and internal states of the model
at arbitrary points of model execution can also be used by the strategy and by the other components. The test
generation framework supplies a compiler for the development of SUT models and implements the measurement of
model coverage, the rule engine and the goal engine.
\par Use of the framework in different real life projects demonstrated its high efficiency, excellent
scalability and good performance resulting in a reduction of testing costs and improved maintainability and
changeability of the created test suites.